\documentstyle[12pt]{article}

\begin{document}

\title{Production of a $Q^2\bar{Q}^2$ state in $J/\psi\rightarrow
\gamma\omega\phi$}
\author{Bing An Li\\
Department of Physics and Astronomy, University of Kentucky\\
Lexington, KY 40506, USA}

\maketitle

\begin{abstract}

The resonance X(1810) discovered in
$J/\psi\rightarrow \gamma\omega\phi$ by BESII is considered as 
a candidate of $0^{++}$ $Q^2\bar{Q}^2$ state.
This model predicts
that $X\rightarrow\omega\phi, K^* K^*$ are the two dominant decay channels and
$X\rightarrow KK, \eta\eta, \eta\eta'$ are suppressed. The cross sections of $\gamma\gamma\rightarrow
X\rightarrow \omega\phi, K^* K^*$ are estimated.
\end{abstract}
\newpage

Hadron spectrum is always one of important topics of particle physics and the test ground of 
nonperturbative QCD. In recent few years both experimental and theoretical studies of
hadron spectroscopy are very active. There are many new discoveries. Most recently BESII has reported
a new resonance near the threshold of $\omega\phi$ in $J/\psi\rightarrow \gamma\omega\phi$[1].
Preliminary fit with a s-wave Breit-Wigner formula leads to
\begin{eqnarray}
\lefteqn{m=1810^{+19}_{-26}\pm13 MeV,\;\;\;
\Gamma=105\pm20\pm11 MeV,}\nonumber \\
&&BR(J/\psi\rightarrow \gamma X(1810), X(1810)\rightarrow\omega\phi)=(2.61\pm0.27\pm0.45)\times
10^{-4}.
\end{eqnarray}
$0^{++}$ are the possible quantum numbers. This resonance is labeled as X(1810) in this letter.

In the quark model the couplings between $u\bar{u}+d\bar{d}$ meson and KK and $K^*K^*$
are OZI allowed and couplings with $\phi\phi, K^* K^*, KK$ are allowed for a $s\bar{s}$ meson too.
If X(1810) is a $q\bar{q}$ meson, then $X(1810)\rightarrow\omega\phi$ is suppressed by the OZI rule
and the coupling between X(1810) and $\omega\phi$ should be very weak. On the other hand,
the phase space of the decay $X(1810)\rightarrow\omega\phi$ is very small. Putting these two
factors together, a very small branching ratio $BR(J/\psi\rightarrow \gamma X(1810), X(1810)
\rightarrow\omega\phi)$ should be expected for a $q\bar{q}(1810)$ state. $J/\psi\rightarrow\gamma
f_0(1710)\rightarrow\gamma KK$ is an example of production of a $0^{++}$ meson in $J/\psi$ radiative decays,
which is an OZI allowed and  
whose branching ratio is measured to be
$(8.5^{+1.2}_{-0.9})\times10^{-4}$[2]. Comparing with this decay and taking very small phase space 
of $X\rightarrow\omega\phi$ into account, the branching ratio of $J/\psi\rightarrow \gamma X(1810), 
X(1810)\rightarrow\omega\phi$
is not small(1). It order to show how weak of the coupling of an OZI suppressed process the well known
processes $\phi\rightarrow KK,\rho\pi$ are taken as examples. It is known that $\phi\rightarrow KK$ channels 
are OZI 
allowed and $\phi\rightarrow\rho\pi$ are OZI suppressed. The general Lagrangian of both processes are
\begin{eqnarray}
{\cal L}_{\phi KK}&=&g_1\phi_\mu\{K^+\partial_\mu K^- -K^-\partial_\mu K^+ +K^0\partial_\mu \bar{K}^0 
-\bar{K}^0\partial_\mu K^0\},\nonumber \\
{\cal L}_{\phi\rho\pi}&=&g_2 {2\over f_\pi}\epsilon^{\mu\nu\alpha\beta}\partial_\mu\phi_\nu\partial_\alpha
\rho^i_\beta\pi^i.
\end{eqnarray}
${\cal L}_{\phi\rho\pi}$ is the ${\cal L}_{\omega\rho\pi}$ which is Wess-Zumino-Witten anomaly[3].
$f_\pi$ is the pion decay constant and $f_\pi=0.184$ GeV is taken. 
The couplings $g_1$ and $g_2$ are determined by the decay widths 
\[g^2_1=20.5,\;\;\;g^2_2=0.0125.\]
The strength of the OZI suppressed channel is three order of magnitude weaker than the 
strength of the OZI allowed. Therefore, if X(1810) is a $q\bar{q}$ meson much smaller 
$BR(J/\psi\rightarrow\gamma X, X\rightarrow\omega\phi)$ than the data should be expected
and it is very difficult to fit X(1810) into the spectrum of ordinary $q\bar{q}$ mesons.

In Ref.[4] the spectrum and the properties of $Q^2\bar{Q}^2$ states have been studied in the MIT bag model.
Some of the $Q^2\bar{Q}^2$ states decay to vector-vector mesons
dominantly by "fall apart" and their masses are at the threshold of the two vector mesons.
The authors of Ref.[5] presents a lattice gauge calculation which shows that the light scalar 
mesons are $Q^2\bar{Q}^2$ states rather than $q\bar{q}$.

In Ref.[6] the $Q^2\bar{Q}^2$ states have been used to study $\gamma\gamma\rightarrow
\rho\rho$. 
Both $0^{++}$ and $2^{++}$ four quark states of \(I=0, 2\),
whose masses are in the range of the mass of $\rho\rho$[4], 
contribute to $\gamma\gamma\rightarrow\rho\rho$. For $\rho^0\rho^0$
the interference between $Q^2\bar{Q}^2$ states of $I=0,2$ is constructive and for $\rho^+\rho^-$
the interference is destructive. The data of $\sigma(\rho^0\rho^0)>>\sigma(\rho^+\rho^-)$ are explained naturally.
The large cross section of $\gamma\gamma\rightarrow
\rho^0\rho^0$ and the position of the peak are explained pretty well in the model of four quark states.
In Ref.[7]
the model of $2^{++}$ $Q^2\bar{Q}^2$ states of $I=0,1$ successfully explains why 
$\sigma(\gamma\gamma\rightarrow K^{*0}\bar{K}^{*0})>>
\sigma(\gamma\gamma\rightarrow K^{*+}K^{*-})$.  
In Ref.[8] the productions of $Q^2\bar{Q}^2$ in 
$J/\psi\rightarrow\gamma VV$ and hadron collisions have been studied.

In this letter 
the possibility of X(1810) as a $Q^2\bar{Q}^2$ state is investigated.
According to Ref.[4], there is a
$C^s(\underline{9}^*)$ $0^+$ $Q^2\bar{Q}^2$ state whose mass has been computed
to be 1.8 GeV[4]. The flavor wave function of $C^s(\underline{9}^*)$ is
\begin{equation}
C^s(\underline{9^*})={1\over\sqrt{2}}K\bar{K}+{1\over\sqrt{2}}\eta_0\eta_s,
\end{equation}
where \(\eta_0={1\over\sqrt{2}}(u\bar{u}+d\bar{d}), \eta_s=s\bar{s}\), the mesons in Eq.(3)
could be either pseudoscalars or vectors.
The color wave function of $Q^2\bar{Q}^2$ state consists of color octet-
color octet and color singlet-color singlet two parts. The recoupling coefficients of $C^s(\underline{9}^*)$
are shown in Table I[8d].   
\begin{table}[h]
\begin{center}
\caption {Table I Recoupling Coefficients}
\begin{tabular}{|c|c|c|c|c|} \hline
    & PP  &  VV & $\underline{P}\cdot\underline{P}$&$ \underline{V}\cdot\underline{V}$  \\ \hline
$9^*$ & -0.177    & 0.644 &0.623 &0.407 \\ \hline         
\end{tabular}
\end{center}
\end{table}
The color octet is indicated by underline.

The decays of a four quark state are through a mechanism called "fall apart"[4]. This
mechanism has been used in previous studies[5,6,7,8]. According to the mechanism of "fall apart",
$\omega\phi, K^* K^*, KK, \eta\eta, \eta\eta'$ are allowed decay channels for $C^s(\underline{9}^*)$ and
the amplitude of a decay of $C^s(\underline{9}^*)$ is proportional to the corresponding recoupling 
coefficient(see Tab.I) of VV or PP and is at s-wave.
The matrix elements of $X(1810)\rightarrow\omega\phi, K^* K^*$ are written as
\begin{eqnarray}
<V_1 V_2|T|X>&=&{1\over\sqrt{2}}0.644 a \epsilon^{\lambda_1}\cdot\epsilon^{\lambda_2},
\end{eqnarray}
where a is taken as a constant in range of X(1810), $\epsilon^{\lambda_1}$ and $\epsilon^{\lambda_2}$
are the polarization vectors of the two vector mesons respectively. For $X\rightarrow KK, \eta\eta,
\eta\eta'$ 
\begin{eqnarray}
<KK|T|X>&=&-{1\over\sqrt{2}}0.177 a,\nonumber \\
<\eta\eta|T|X>&=&-{1\over\sqrt{2}}0.177 a(-{\sqrt{2}\over3}cos2\theta-{1\over6}sin2\theta),\nonumber \\
<\eta\eta'|T|X>&=&-{1\over\sqrt{2}}0.177 a(-{1\over3}cos2\theta+{2\sqrt{2}\over3}sin2\theta),
\end{eqnarray}
where $\theta$ is the mixing angle of $\eta-\eta'$ and $\theta=-11^0$ is taken in this paper.
The decay widths of various channels are obtained
\begin{eqnarray}
\Gamma(X\rightarrow\omega\phi)&=&({1\over\sqrt{2}}0.644 a)^2\frac{k_1}{8\pi q^2}
\{2+{1\over4m^2_\omega m^2_\phi}(q^2-m^2_\omega-m^2_\phi)^2\},\nonumber \\
k^2_1&=&{1\over4q^2}\{(q^2-m^2_\omega-m^2_\phi)^2-4m^2_\omega m^2_\phi\},\nonumber \\
\Gamma(X\rightarrow K^* K^*)&=&({1\over\sqrt{2}}0.644 a)^2\frac{k_2}{8\pi q^2}
\{2+{1\over4m^4_{K^*}}(q^2-2m^2_{K^*})^2\},\;\;\;
k^2_2={q^2\over4}-m^2_{K^*},\nonumber \\
\Gamma(X\rightarrow KK)&=&({1\over\sqrt{2}}0.177 a)^2\frac{k_3}{8\pi q^2},\;\;\;
k^2_3={q^2\over4}-m^2_K,\nonumber \\
\Gamma(X\rightarrow \eta\eta)&=&({1\over\sqrt{2}}0.177 a)^2\frac{k_4}{4\pi q^2}
(-{\sqrt{2}\over3}cos2\theta-{1\over6}sin2\theta)^2,\;\;\;
k^2_4={q^2\over4}-m^2_\eta,\nonumber \\
\Gamma(X\rightarrow \eta\eta')&=&({1\over\sqrt{2}}0.177 a)^2\frac{k_5}{8\pi q^2}
(-{1\over3}cos2\theta+{2\sqrt{2}\over3}sin2\theta)^2,\nonumber \\
k^2_5&=&{1\over4q^2}\{(q^2-m^2_\eta-m^2_{\eta'})^2-4m^2_\eta m^2_{\eta'}\},
\end{eqnarray}
where q is the momentum of X(1810). Adding up all the widths, the total width of X(1810) is obtained.
Using the value of $\Gamma$(1), the parameter a is determined to be $7.1(1\pm0.11)$GeV.
The ratios of the widths are determined
\begin{eqnarray}
\frac{\Gamma(X\rightarrow K^* K^*)}{\Gamma(X\rightarrow\omega\phi)}=1.83,\;\;\;
\frac{\Gamma(X\rightarrow K K)}{\Gamma(X\rightarrow\omega\phi)}=0.22,\;\;\;
\frac{\Gamma(X\rightarrow \eta\eta)}{\Gamma(X\rightarrow\omega\phi)}=0.059,\;\;\;
\frac{\Gamma(X\rightarrow \eta\eta')}{\Gamma(X\rightarrow\omega\phi)}=0.062.
\end{eqnarray}
These ratios are independent of parameter a.
The model of $Q^2\bar{Q}^2$ predicts that
$K^* K^*$ and $\omega\phi$ are the two dominant decay channels, 
the rate of KK channel is 
less than $\omega\phi$ by a factor 4.5, and  
the decay rates of $\eta\eta, \eta\eta'$ are very small. The dominance of the VV channels is resulted in
three factors : the ratio of recoupling coefficients $(0.644/0.177)^2$, s-wave, and three
independent directions of polarizations of vector meson. Of course, for KK channel the phase 
space is larger. The effects of the mixing angle between $\eta-\eta'$ make the decay rates of 
$\eta\eta, \eta\eta'$ channels smaller. 

In QCD $J/\psi$ radiative decays are described as
$J/\psi\rightarrow\gamma gg, gg\rightarrow hadrons$. 
The $0^{++}$ four quark state $C^s(\underline{9^*})$ can via the 
$\underline{\omega}\cdot\underline{\phi}$ 
component(see Tab.I) 
couple to two gluons and decays to $\omega\phi$. 
The process $J/\psi\rightarrow\gamma\omega\phi$ is described as
$J/\psi\rightarrow\gamma gg, gg(\underline{V}\underline{V})\rightarrow Q^2\bar{Q}^2\rightarrow
\omega\phi$.
As a matter of fact, the process $gg\rightarrow Q^2\bar{Q}^2\rightarrow VV$ has been used in the
studies of the productions of $Q^2\bar{Q}^2$ states in hadron collisions and $J/\psi$ radiative decays[8].
$gg(\underline{V}\underline{V})\rightarrow
Q^2\bar{Q}^2$ is at $O(\alpha_s)$ and $Q^2\bar{Q}^2\rightarrow VV$ at $O(\alpha^0_s)$. Therefore,
$J/\psi\rightarrow\gamma gg, gg(\underline{V}\underline{V})\rightarrow 
Q^2\bar{Q}^2\rightarrow VV$ is at the same order as $J/\psi\rightarrow\gamma+hadrons(q\bar{q})$.
If X(1810) is a $Q^2\bar{Q}^2$ state there is no suppression 
for $J/\psi\rightarrow\gamma X(1810), X(1810)\rightarrow\omega\phi$.  

At lower energies the electromagnetic interactions of hadrons are determined by 
the Vector Meson Dominance(VMD)[9]. The substitutions of the VMD for $\omega$ and $\phi$ mesons are
expressed as  
\begin{equation}
\omega\rightarrow{1\over6}egA,\;\;\phi\rightarrow-{\sqrt{2}\over6}egA,
\end{equation}
where $g=0.39$ is determined in Ref.[10]. Using Eqs.(8), a vector meson can be replaced by a photon
and corresponding Lagrangian of electromagnetic interactions of hadrons is obtained. 
It is similar to the VMD(8) the substitutions
between a gluon and color octet vectors are obtained in Ref.[8](different notations are used in this paper)
\begin{equation}
\underline{\omega}^a\rightarrow{1\over2}g_s g_{\underline{v}}g^a,\;\;\;
\underline{\phi}^a\rightarrow{\sqrt{2}\over2}g_s g_{\underline{v}}g^a,
\end{equation}
where $g_{\underline{V}}^2={2\over3}g^2$ is determined and $g_s$ is the coupling constant of QCD.
Using Eqs.(9), 
\begin{equation}
{\cal L}_{X gg}=0.407a{1\over4}g^2_s g^2_{\underline{v}} X g^a_\mu g^a_\mu
\end{equation}
is determined. 
In order to satisfy gauge invariance in the momentum picture the tensor $g_{\mu\nu}$ 
of $g_{\mu\nu}g^a_\mu g^a_\nu$ is replaced by $(-g_{\mu\sigma}+\frac{k_{1\mu}k_{1\sigma}}{k^2_1})
(-g_{\nu\sigma}+\frac{k_{2\mu}k_{1\sigma}}{k^2_1})$, where $k_{1,2}$ are the momentum of two 
gluons respectively. Instead direct calculation of the matrix element of $J/\psi\rightarrow\gamma gg$
the approach of effective Lagrangian is exploited in this paper. 
The simplest effective Lagrangian of $J/\psi\rightarrow\gamma gg$ is constructed as
\begin{equation}
{\cal L}_{J\gamma gg}={eA\over m^4_{J/\psi}}(\partial_\mu J_\nu-\partial_\nu J_\mu)
(\partial_\mu A_\nu-\partial_\nu A_\mu)(\partial_\lambda g^a_\sigma-\partial_\sigma
g^a_\lambda)(\partial_\lambda g^a_\sigma-\partial_\sigma
g^a_\lambda),
\end{equation}
where A is taken as a parameter. This effective Lagrangian is gauge invariant.

Using Eqs.(10,11), the decay rate of $J/\psi\rightarrow\gamma X(1810), X(1810)\rightarrow f$ is derived 
\begin{eqnarray}
\frac{d\Gamma}{dq^2}&=&\alpha{96\over\pi}{A^2\over m^{11}_{J/\psi}}(0.407a {1\over4}g^2_s 
g^2_{\underline{V}})^2\sqrt{q^2}
(m^2_{J/\psi}-q^2)^3 f^2(q^2)\frac{\Gamma_{X\rightarrow f}(q^2)}{(q^2-m^2_X)^2+q^2\Gamma^2},\nonumber \\
f(q^2)&=&-\frac{1}{(2\pi)^3}\int^{q^2}_0 dx (q^2+2x)(q^2-x)\int^{{\pi\over2}}_0 d\phi\frac{sin^2\phi}
{(q^2-x)^2+4q^2 xcos^2\phi},
\end{eqnarray}
where the function $f(q^2)$ represents the effects of gluons in the process $J/\psi\rightarrow\gamma gg,
gg\rightarrow X$ and $\Gamma_{X\rightarrow f}$ is shown in Eqs.(6).
\(\alpha_s={g^2_s\over4\pi}=0.3\) is taken.
The ratios of the branching ratios of various channels of 
$J/\psi\rightarrow\gamma X, X\rightarrow f$ are obtained from Eq.(12)
\begin{eqnarray}
\frac{BR(K^* K^*)}{BR(\omega\phi)}&=&1.44,\;\;\frac{BR(K K)}{BR(\omega\phi)}=0.18,\nonumber\\
\frac{BR(\eta\eta)}{BR(\omega\phi)}&=&4.8\times10^{-2},\;\;\frac{BR(\eta\eta')}{BR(\omega\phi)}
=7.5\times 10^{-2},\nonumber \\
BR(K^* K^*)&=&3.76(1\pm0.2)\times10^{-4},\nonumber \\
BR(KK)&=&0.47(1\pm0.2)\times10^{-4},\;\;BR(\eta\eta)=1.26(1\pm0.2)\times10^{-5},\nonumber\\
BR(\eta\eta')&=&1.97(1\pm0.2)\times10^{-5}.
\end{eqnarray}
The ratios are independent of parameters A and a.
The upper limit of the integrals over $q^2$ is taken as $(1.81+0.105)^2 GeV^2$ and the lower limits are
$(m_\omega+m_\phi)^2$, $4m^2_{K^*}$, and $(1.81-0.105)^2$ for $\omega\phi$, $K^* K^*$, 
and PP channels respectively. Because of 
the dependence on $q^2$ and the choices of the lower limits of $\frac{d\Gamma}{dq^2}
(J/\psi\rightarrow\gamma X, 
X\rightarrow f)$ the ratios shown 
in Eqs.(13) are not the same as shown in Eqs.(7). However, the same predictions are made by the model of 
$Q^2\bar{Q}^2$ 
that $K^* K^*$ and $\omega\phi$
are two dominant channels and the KK, $\eta\eta$, $\eta\eta'$ channels are suppressed.
The branching ratio of $J/\psi\rightarrow\gamma K^* K^*$ has been measured[11].
The data were fitted by simple Breit-Wigner resonances of $0^{-+}, 2^{++}$, 
and $2^{-+}$. A broad $0^-$ at $1800\pm100$ MeV with width at $500\pm200$ MeV is revealed. 
The branching ratio of $J/\psi\rightarrow\gamma (K^* K^*)_{0^+}$ predicted in this paper(13) is less than 
the branching ratios of $0^{-+}, 2^{++}$, and $2^{-+}$[11]. Therefore, a smaller $0^{++}$ component is not
ruled out. 

From Eq.(12) 
\begin{equation}
BR(J/\psi\rightarrow\gamma\omega\phi)=1.06\times10^{-3}A^2 GeV.
\end{equation}
Inputting the data(1), \(A^2=0.25(1\pm0.2)\) is determined in the region of the mass of X(1810). 
In order to justify that ${\cal L}_{J\gamma gg}$(11) is a reasonable effective Lagrangian the decay rate of
$J/\psi\rightarrow\gamma (gg)_{0^+}$ is computed 
\begin{eqnarray}
BR(J/\psi\rightarrow\gamma (gg)_{0^+})&=&0.016 A^2=4.0(1\pm0.2)\times10^{-3}.
\end{eqnarray}
The invariant mass of $(gg)_{0^{+}}$ is taken in the range of $1.81\pm0.05$GeV. Comparing with the
branching ratios of other $J/\psi$ radiative decays, the value of $BR(J/\psi\rightarrow\gamma (gg)_{0^+})$
is acceptable.

As a $Q^2\bar{Q}^2$ state X(1810) can be produced in $\gamma\gamma$ collisions. Using the VMD(9),
\begin{equation}
{\cal L}_{X\gamma\gamma}=-{1\over36}e^2 g^2 0.644a XA_\mu A_\mu
\end{equation}
is obtained. The problem of gauge invariance of Eq.(16) can be solved by the same way as in the case
of Eq.(10). As a matter of fact, when the two photons are on mass shell the extra factor for gauge 
invariance doesn't contribute. From Eqs.(4,16) the cross section of $\gamma\gamma\rightarrow X\rightarrow
 f$ is derived
\begin{equation}
\sigma=32\pi^2\alpha^2(0.0179ag^2)^2{1\over \sqrt{q^2}}\frac{\Gamma(q^2)_{X\rightarrow f}}
{(q^2-m^2_X)^2+q^2\Gamma^2}.
\end{equation}
At the peak \(q^2=m^2_X\) 
\begin{eqnarray}
\sigma(\gamma\gamma\rightarrow X\rightarrow\omega\phi)&=&1.24 nb,\;\;\;
\sigma(\gamma\gamma\rightarrow X\rightarrow K^* K^*)=2.24 nb.
\end{eqnarray}
These results(18) are parameter independent. In Ref.[7] the states of $2^{++}$ $Q^2\bar{Q}^2$ of $I=0,1$
have been used to study $\gamma\gamma\rightarrow K^* K^*$ in the range of 1.7-2.7GeV. This study(18)
shows that the contribution of $0^{++}$ $Q^2\bar{Q}^2$ of $I=0$ to $\gamma\gamma\rightarrow K^* K^*$
is small. However, the measurements of $\sigma(\gamma\gamma\rightarrow\omega\phi)$ is very interesting.

On the other hand, if X(1810) is a $C^s(\underline{9}^*)$ state this $\omega\phi$ resonance can be produced
in other processes without OZI suppression. The Tab.I shows X(1810)($C^s(\underline{9}^*)$) has stronger
couplings with both $\omega\phi$ and $K^* K^*$. Therefore, it can be produced 
not only in $gg\rightarrow X(1810)\rightarrow f$ but in the processes 
$\omega\phi(K^* K^*)\rightarrow X(1810)\rightarrow f$ and $\gamma\omega(\phi)\rightarrow X(1810)\rightarrow
f$ too. 
As a matter of fact,
the amplitude of the process $\gamma\gamma\rightarrow X(1810)\rightarrow\omega\phi(K^* K^*)$ 
studied in this paper is obtained via the VMD from $\omega\phi\rightarrow X(1810)
\rightarrow\omega\phi(K^* K^*)$. 
$\omega\phi$ and $K^* K^*$ of $I(J^{PC})=0(0^{++})$ can be produced in hadron collisions, photo-productions, 
decays of B mesons, and $J/\psi$ hadronic decays. 
The study of these processes are beyond 
the scope of this letter. 

It is interesting to notice that there is a $C^s(\underline{9})$ of $2^{++}$ state whose mass is 1.95GeV[4].
This state decays to $\omega\phi$ and $K^* K^*$ only. The contributions of $C^s(\underline{9})$ in various
processes have been studied[7,8]. It is possible that this state appears in the tail of BESII's data.

In summary, because the decay mode $q\bar{q}\rightarrow\omega\phi$ is suppressed by the OZI rule
the decay $X\rightarrow\omega\phi$ with not small branching ratio is very difficult to be 
understood if X(1810) is a $q\bar{q}$ meson. On the other hand, the mass, quantum number, and the decay 
$J/\psi\rightarrow\gamma X(1810), X(1810)\rightarrow\omega\phi$ can be explained by assigning X(1810) to  
a $Q^2\bar{Q}^2$ state, $C^s(\underline{9}^*)$[4]. There is no suppression for $C^s(\underline{9}^*)
\rightarrow\omega\phi$. 
The model of $Q^2\bar{Q}^2$ predicts: 1)besides the $\omega\phi$ channel there is another
dominant decay channel, $K^* K^*$; 2)the decay channels
KK, $\eta\eta$, $\eta\eta'$ channels are suppressed; 3)the cross sections
of the productions of X(1810) in two photons collisions are estimated. 
4)as $C^s(\underline{9}^*)$ X(1810) can be produced either in two gluon fusion 
$gg\rightarrow X(1810)\rightarrow \omega\phi(K^* K^*)$ or the fusion of two vector mesons, 
$VV(\omega\phi, K^* K^*)\rightarrow X(1810)\rightarrow\omega\phi(K^* K^*)$. 

The author would like to thank S.Jin, W.G Li and 
X.Y.Shen for discussions.

\end{document}